\documentclass[aps,prl,twocolumn,superscriptaddress,showpacs, nobalancelastpage, nofootinbib]{revtex4-1}
\usepackage{graphicx}
\usepackage{subfigure} 
\usepackage{float} 
\usepackage[utf8]{inputenc}
\usepackage{latexsym}
\usepackage{amsmath}
\usepackage{txfonts}
\usepackage{helvet}
\usepackage{braket}
\usepackage{amscd}
\usepackage[official]{eurosym}
\usepackage{longtable}
\usepackage{amsfonts}
\usepackage{siunitx}
\usepackage{booktabs}
\usepackage{amssymb}  
\usepackage[]{units}

\begin{document}

\title{A Method for Measurement of Spin-Spin Couplings with sub-mHz Precision Using Zero- to Ultralow-Field Nuclear Magnetic Resonance} 
\author{A. Wilzewski}
\affiliation{Johannes Gutenberg-Universit{\"a}t, 55099 Mainz, Germany}
\author{S. Afach}
\affiliation{Johannes Gutenberg-Universit{\"a}t, 55099 Mainz, Germany}
\author{J. W. Blanchard}
\email{blanchard@uni-mainz.de}
\affiliation{Helmholtz-Institut Mainz, 55099 Mainz, Germany}
\author{D. Budker}
\affiliation{Johannes Gutenberg-Universit{\"a}t, 55099 Mainz, Germany}
\affiliation{Helmholtz-Institut Mainz, 55099 Mainz, Germany}
\affiliation{University of California at Berkeley, Berkeley, CA}
\affiliation{Nuclear Science Division, Lawrence Berkeley National Laboratory, Berkeley, CA}

\date{\today}

\begin{abstract} 
We present a method which allows for the extraction of physical quantities directly from zero- to ultralow-field nuclear magnetic resonance (ZULF NMR) data. 
A numerical density matrix evolution is used to simulate ZULF NMR spectra of several molecules in order to fit experimental data. 
The method is utilized to determine the indirect spin-spin couplings ($J$-couplings) in these, which is achieved with precision of $10^{-2}$--$10^{-4}$ Hz.
The simulated and measured spectra are compared to earlier research.
Agreement and precision improvement for most of the $J$-coupling estimates are achieved.
The availability of an efficient, flexible fitting method for ZULF NMR enables a new generation of precision-measurement experiments for spin-dependent interactions and physics beyond the Standard Model.

\end{abstract}

\maketitle

\section{Introduction}
Nuclear magnetic resonance (NMR) 
has many applications, including precision measurement of physical constants, chemical detection and analysis, and bimolecular structure elucidation.
In conventional NMR, large magnetic fields are used in order to enhance chemical-shift resolution and improve the sensitivity of inductive detection, as well as to increase signals due to higher polarization.
Furthermore, working at high-field also serves to ``truncate'' second-order effects arising from nuclear spin interactions that do not commute with the Zeeman Hamiltonian.
Zero- to ultralow-field (ZULF) NMR is a recently developed alternative method that does not utilize strong magnetic fields \cite{Blanchard2013,Theis2013,Butler2013,Ledbetter2013Exp,Blanchard2016}, at least not for encoding and detection.
Due to the high absolute field homogeneity and the absence of some relaxation pathways such as those related to chemical-shift anisotropy, ZULF NMR frequently achieves narrow resonance linewidths in the order of tens of \si{\milli\hertz}, allowing for precise measurement of spin-spin interactions.
Additionally, the absence of truncation by a large applied magnetic field means that ZULF NMR is capable of measuring spin-dependent interactions that do not commute with the Zeeman Hamiltonian \cite{Blanchard2015}, which is not generally possible in conventional NMR experiments.
%

ZULF NMR operates in the regime where the Zeeman Hamiltonian vanishes or can be treated as a perturbation compared to the internal interactions among the nuclei of the molecule.
For isotropic liquids, the direct dipole-dipole couplings are averaged due to the random motion of the molecules.
The main interactions remaining are the electron-mediated indirect spin-spin coupling, $J$-couplings, of the form $J \textbf{I}_1\cdot\textbf{I}_2$, between two nuclear spins $\textbf{I}_1$ and $\textbf{I}_2$.
These couplings are dependent on the geometry and electronic structure of the molecule. This makes $J$-spectroscopy a source on information for chemical analysis and fingerprinting.
Moreover, $J$-couplings are a source of information of spin topology and torsion and bond angles. 
For instance, a structural analysis of several benzene derivatives is discussed in Ref.\cite{Blanchard2013}.
Additional information can be provided by applying small magnetic fields \cite{PhysRevLett.107.107601} or by reintroducing molecular alignment in, for example, stretched polymer gels \cite{Blanchard2015}.

Recently, $J$-couplings have also drawn the attention of physicists searching for anomalous spin-dependent forces \cite{Ledbetter2013Axion} arising from axion-like particles \cite{Axion1984}, which are possible dark-matter candidates \cite{DarkmatterAxion}.
For example, in Ref.~\cite{Ledbetter2013Axion}, the authors were able to set new constraints on the coupling constants between nucleons mediated by exchange of pseudoscalar (axion-like) and axial-vector bosons by comparing precision measurements and calculations of the $J$-coupling in hydrogen deuteride.
More recent experiments have hinted that there may be a significant difference between measured and predicted values for the HD $J$-coupling \cite{Neronov2015}.
However, while the authors of Ref.~\cite{Garbacz2016} also measure slight discrepancies between experiment and theory for the couplings in hydrogen deuteride (HD), hydrogen tritide (HT), and deuterium tritide (DT), they suggest that the effect may be associated with the absence of nonadiabatic corrections in the theory.
Along with improvements to calculations, additional precise $J$-coupling measurements will be valuable to these efforts.
Furthermore, it may be possible to search for other exotic physics mediated by new particles by fitting ZULF NMR spectra using various model Hamiltonians that include exotic interactions such as those identified in Ref.~\cite{Dobrescu2006}.
Performing such fits is required for high-precision measurement of $J$-couplings and/or testing model Hamiltonians. 
This necessitates the development of efficient code to achieve this task.

In this paper, we introduce a method for measurement of $J$-couplings that uses ZULF NMR data and provides high precision below spectral linewidths.
As a proof of principle, we provide measured $J$-couplings for several molecules with precision comparable to or better than that currently available in the literature.
This method can be readily modified to include any additional spin interactions. 
The zero-field data of Refs. \cite{Blanchard2013,Butler2013,Theis2013} were reanalyzed, yielding sub-mHz precision for some of the $J$-couplings as discussed below.
\begin{figure}
\includegraphics[width=\columnwidth]{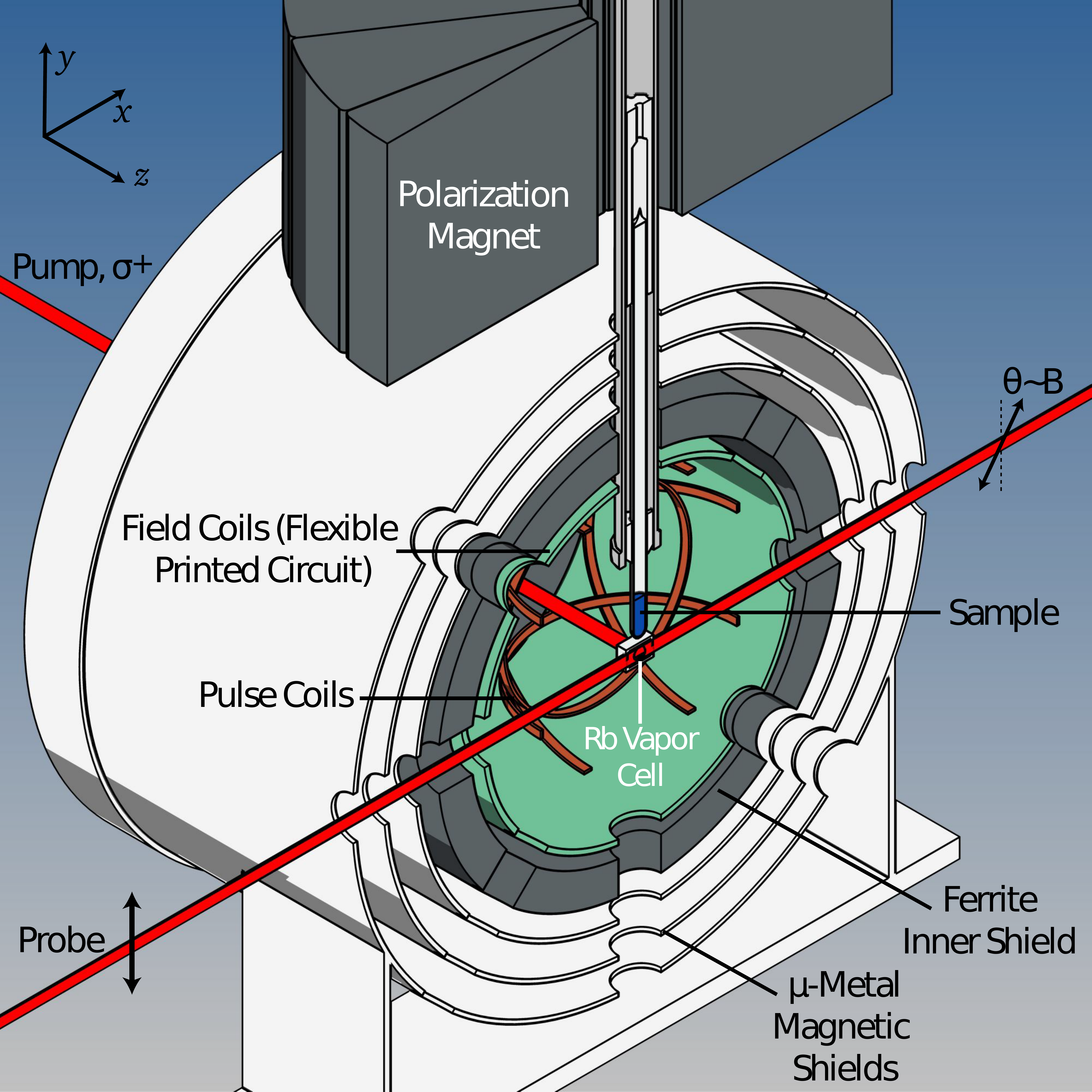}
\caption{Schematic of the ZULF apparatus: magnetic field in $z$-direction is measured by a ${}^{87}$Rb alkali-vapor optical atomic magnetometer. Magnetic field is probed by measuring polarization rotation of a linearly polarized beam. The sample is shuttled pneumatically into the magnetically shielded detection region after being polarized.} 
\label{schematic}
\end{figure}
\section{ZULF NMR Experiments}

The experiments are discussed in Refs. \cite{Blanchard2013,Blanchard2016}. 
A schematic of the zero-field NMR apparatus is shown in Fig.~\ref{schematic}.
The sample is polarized in a $2\,\unit{T}$ permanent Halbach magnet and is then pneumatically shuttled to the detection region.
A ${}^{87}$Rb alkali-vapor optical atomic magnetometer operating in the spin-exchange relaxation-free (SERF) regime \cite{SERF2002} is used to measure the evolution of the sample's nuclear spin magnetization.
The alkali vapor cell of the optical magnetometer is surrounded with two sets of three orthogonal coils: one set is used for the application of magnetic field pulses, and the other for cancellation of residual static magnetic fields.
The atoms in the cell are pumped with a circularly polarized laser light tuned to the D1 transition, propagating along the +$z$-direction.
The magnetic field is probed by optical rotation (OR) of a second frequency-detuned beam propagating along +$x$ direction. 
With this configuration, the magnetometer is sensitive to magnetic fields in the $y$-direction.
After shuttling the sample to the detection region, a pulse (for instance, with area $B_{x} t=\pi/\gamma_C$, for ${}^{13}$C-labeled molecules in $x$-direction, where $\gamma_C$ is the gyromagnetic ratio of ${}^{13}$C) induces coherence among nuclear spins in the molecule.
The evolution of this system is then measured with the magnetometer.
The signal is processed and Fourier-transformed to create the spectrum that is discussed in later sections of this work.
\begin{figure*}
\includegraphics[width=0.8\textwidth]{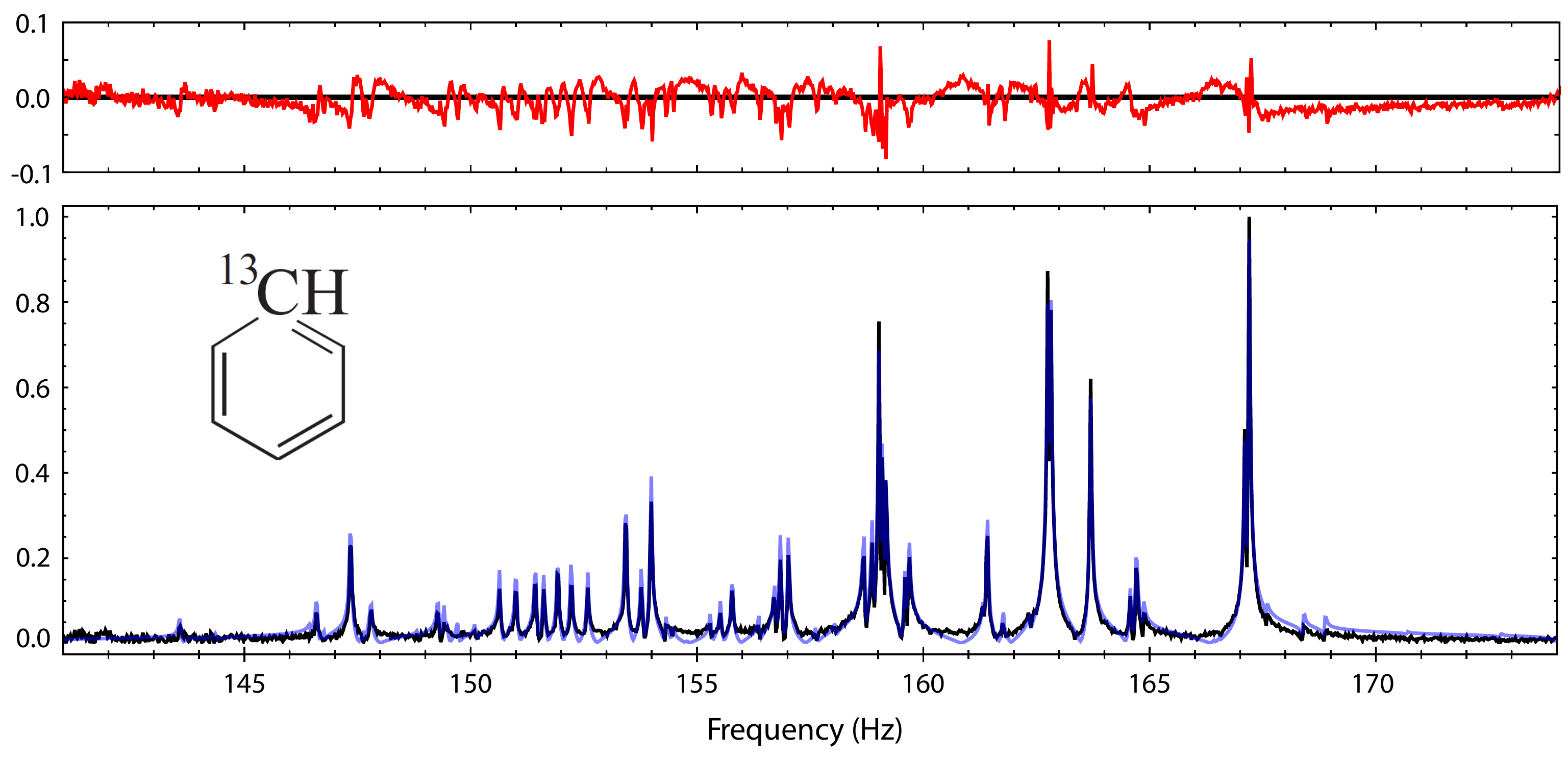}
  \caption{Measured zero-field NMR spectrum (black) of benzene-${}^{13}$C$_1$ in the vicinity of ${}^1J_{\text{CH}}$ with the fitted spectrum (blue) overlaid. The noisy region between 50 Hz and 80 Hz as well as 2 Hz intervals around overtones of the 60 Hz line noise were excluded from the fit. Fit residuals are shown in red above.}
\label{Fig:benzene}
\end{figure*}

\section{Fitting Method}
A density-matrix simulation was developed to model the experiment.
The simulation starts by assuming a density matrix corresponding to thermal equilibrium in a magnetic field $B_y$
\begin{equation}
\rho_{0}=\frac{e^{-\beta H_{pol}}}{\text{Tr}\left(e^{-\beta H_{pol}}\right)},
\end{equation}
where $\beta=(k_B T)^{-1}$ with Boltzmann constant $k_B$ and temperature $T$, and $H_{pol} = - B_y \sum_j \gamma_j I_{y,j}$ is the Zeeman Hamiltonian during the polarization step with gyromagnetic ratio $\gamma_j$ of nuclear spin $\textbf{I}_j$.
For all future steps, the polarizing magnetic field $B_y$ is removed.
Then, an instantaneous magnetic field pulse (in this case a $\pi$-pulse in $x$-direction on the ${}^{13}$C) is applied,
\begin{equation}
\rho \left(t=0\right)=U^\dagger \rho_{0} \, U,
\end{equation}
with $U = \exp \left({-\nicefrac{i}{\hbar} \cdot \sum_j \gamma_j I_{x,j} \cdot \nicefrac{\pi}{\gamma_C}}\right)$.
We neglect all other interactions including $J$-coupling during the pulse.
Finally the density matrix is evolved in $T \times f_s$ steps, where $T$ is the measurement period, and $f_s$ is the sample rate of the signal.
The evolution is performed with the operator $e^{i/\hbar \cdot H \cdot t}$, with the Hamiltonian
\begin{equation}
H = \sum_{i,j} J_{ij} \textbf{I}_i \cdot   \textbf{I}_j.
\end{equation}
Every step of the evolution is accompanied by an evaluation of the magnetization \cite{Ledbetter2009} of the system along the quantization axis
\begin{equation}
M_y(t) = \text{Tr}\left(\,\rho(t)\sum_{i} \gamma_i\textbf{I}_{y,i}\right).
\end{equation}
The magnetization signal is Fourier-transformed afterwards to create the simulated spectrum, which is then compared to the experimental spectrum.

The evolution of the magnetization, $M_y(t)$, is a function of the $J$-couplings.
In order to extract the $J$-couplings from the experimental signals, an optimization problem is solved by iterative least-squares fitting that compares the experimental and the simulated spectra.
In addition to the $J$-couplings, the parameters of the least-squares fit include a scaling factor $A$ and a relaxation parameter $\tau$, which are used to produce an ad-hoc exponential decay envelope $A e^{-t/\tau}$ that is multiplied with the signal such that the simulated signal becomes $M_y(t)\Rightarrow A M_y(t) e^{-t/\tau}$.
Using a single exponential decay is justified by the fact that in the data presented here all coherences appear to decohere uniformly, although in general different parts of the density matrix may decay at different rates \cite{LongLivedSpinSinglets}.
Finally the simulated signal is Fourier-transformed and the amplitudes of the simulated and experimental spectra are compared.  
It should be mentioned that the signal simulation is computationally expensive. 
The computational complexity of the problem is given by $O\left(N^3\right)$, where $N$ is the side-length of the density matrix of the system, for example, in the case of $n$ spin-$\nicefrac{1}{2}$ particles, $N=2^n$.
This is because evolving a density matrix involves matrix multiplication, which has, at best, a computational complexity of $O(N^{2.81})$, using the divide-and-conquer algorithm.
By representing the density matrix and the magnetic moment operator in the basis of $H$, we were able to reduce the computational complexity to better than $O\left(N^2\right)$.
This is particularly important for molecular systems with more than five spins, where systematic matrix multiplication becomes prohibitively computationally expensive.
Our program computes the spectra in a time $t=A\cdot (2^n)^2$, where
$A$ is a factor that depends on the computational power of the computer used.
The memory consumption of the program also follows the same formula as a function of the number of spins.
On an office workstation with an Intel 4960X processor overclocked to 4.5 GHz, a spectrum of a system with 9 spins takes 7 seconds to simulate (20000 points), and consumes 450 MB of memory. On such a workstation we expect that the practical limit for a spin system is $\sim$12 spins, due to memory and computational time.

\section{Estimating Uncertainties}
Because no convenient analytic solutions are generally available for the spectra under study, and the topology of $\chi^2$ in the parameter space is commonly complex, the choice of the fitting algorithm requires caution.
Algorithms that actively use derivatives to choose the steepest-decent direction are inefficient, since many extra function evaluations are required for them to work.
Because of this, the Downhill Simplex Algorithm \cite{Simplex} is implemented for the fits presented here.
The simplex algorithm substitutes using derivatives with evaluating different points in the parameter space, and guessing the next step's direction.
Once the minimum $\chi^2_{min}$ is reached the standard deviations of all $p$ parameters are determined via 
\begin{equation}
\sigma_{a_k}^2=\frac{\chi^2_{min}}{n-p}\cdot C_{kk}^{-1},
\end{equation}
where $n = T \times f_s$ is the number of points of the spectrum, $\sigma_{a_k}$ is the standard deviation of parameter $a_k$ and $C_{kk}^{-1}$ is the $k^{th}$ diagonal element of the inverse covariance matrix \cite{Wolberg2006}.\\
%
%
%
%

\section{Results and Discussion}
We now demonstrate how $J$-couplings may be extracted from experimental ZULF NMR spectra of several molecules.
A detailed description of how these spectra can be understood in terms of molecular/electronic structure can be found in Refs. \cite{Blanchard2013,Butler2013,Theis2013,Blanchard2016}.
The purpose of this paper is to focus on the actual extracted values of the $J$-couplings.
Specific noisy parts of the spectrum are excluded from the fits.
For example, due to the fact the data were recorded in the USA, every spectrum shows peaks at 60 Hz and multiples thereof because the mains electrical power in the USA is provided at that frequency.
In order to not fit to those peaks, a 2 Hz interval was cut out of the spectrum around every multiple of 60 Hz (e.g. 179-181~Hz).
Other noisy portions of the spectra deemed unnecessary for $J$-coupling fitting were excluded on a case-by-case basis, as discussed in the following sections.
For example, for the majority of benzene derivatives, the noisy low-frequency part of the spectrum can be ignored because the high-frequency peak-clusters contain enough information to determine the $J$-couplings, as described in \cite{Blanchard2013}.

\subsection{Benzene-${}^{13}$C$_1$}

The experimental ZULF NMR spectrum and fitted model of benzene-${}^{13}$C$_1$ are shown in Fig.~\ref{Fig:benzene}.
The $J$-couplings of benzene have been extensively studied \cite{Chertkov1983,Blanchard2013}, so this case is particularly useful to determine whether the method presented here yields results which are in agreement with former research.
The spectrum from 5 Hz to 300 Hz was fitted while the noisy region from 50 Hz to 80 Hz most likely caused by vibrations in the apparatus was excluded.
This exclusion is also justified by the fact that in this region of the spectrum no peaks generated by the molecular model for benzene-${}^{13}$C$_1$ are present.
In contrast to most of the other molecules the low-frequency part (not visible in Fig. \ref{Fig:benzene}) of the spectrum was maintained partially to fit to all the predicted peaks, even though fits to only the high-frequency cluster lead to similar results.
In the interval 140--175 Hz the residuals are roughly one order of magnitude smaller than the actual peak heights, in contrast to the low-frequency part with larger residuals.
Those also might be caused by vibrations in the apparatus or external sources. 
The $J$-couplings extracted from the fit are listed in Table \ref{Tbl:benzeneResults}, along with the couplings reported in Ref.~\cite{Chertkov1983}.
There is a general agreement between our fitted values and those from Ref.~\cite{Chertkov1983}, though there are a number of couplings for which the disagreement between the two measurements is greater than the quadrature sum of the standard deviations.
Different experimentally measured values for benzene-${}^{13}$C may reasonably be expected due to minor differences in sample preparation procedures (e.g. presence of cosolvents, water absorption, etc.), measurement temperature, etc.
Furthermore, due to benzene's anisotropic magnetic susceptibility, a small degree of molecular alignment will occur in large magnetic fields, giving rise to residual dipolar couplings that may be difficult to distinguish from $J$-couplings, shown for some aromatics in Ref.~\cite{doi:10.1021/ja00149a025}. 

\begin{table}[h]
\begin{center}
\begin{tabular}{c*{1}{c} c c}
$J$-coupling & Fitted value (Hz) & Literature value$^a$ & Difference\\  
\hline
& & \\${}^1J_{\rm CH}$		& 158.363(1)	& 158.354(1) 	&0.009(1)\\ 
${}^2J_{\rm CH}$			& 1.136(8) 		& 1.133(3)		&0.003(9)\\ 
${}^3J_{\rm CH}$			& 7.609(10) 	& 7.607(3)		&0.002(10)\\ 
${}^4J_{\rm CH}$			& -1.285(16) 	& -1.296(4)		&0.011(16)\\
& & 	\\
${}^3J_{\rm HH}(\rm H1,H2)$	& 7.534(9) 		& 7.540(2)		&0.006(9)\\ 
${}^3J_{\rm HH}(\rm H2,H3)$	& 7.543(2) 		& 7.543(2)		&0.000(3)\\ 
& & \\
${}^4J_{\rm HH}(\rm H3,H4)$	& 7.543(1) 		& 7.535(2)		&0.008(2)\\ 
${}^4J_{\rm HH}(\rm H1,H3)$	& 1.381(9) 		& 1.380(2)		&0.001(9)\\ 
${}^4J_{\rm HH}(\rm H2,H4)$	& 1.382(1) 		& 1.377(2)		&0.005(2)\\ 
& & \\
${}^5J_{\rm HH}(\rm H2,H6)$	& 1.384(3) 		& 1.373(4)		&0.011(5)\\
${}^5J_{\rm HH}(\rm H3,H5)$	& 1.387(3) 		& 1.382(4)		&0.005(5)\\
${}^5J_{\rm HH}(\rm H1,H4)$	& 0.658(2) 		& 0.661(3)		&0.003(4)\\
${}^5J_{\rm HH}(\rm H2,H5)$	& 0.660(2) 		& 0.658(2)		&0.002(3)\\
& & \\
\hline
\end{tabular}
\caption{Fit results for benzene-${}^{13}$C$_1$.$^a$From Ref.~\cite{Chertkov1983}.}
\label{Tbl:benzeneResults}
\end{center}
\end{table}

\subsection{Benzaldehyde-$\alpha$-${}^{13}$C$_1$ and Toluene-$\alpha$-${}^{13}$C$_1$}

\begin{figure}
\includegraphics[width=0.49\textwidth,height=130px]{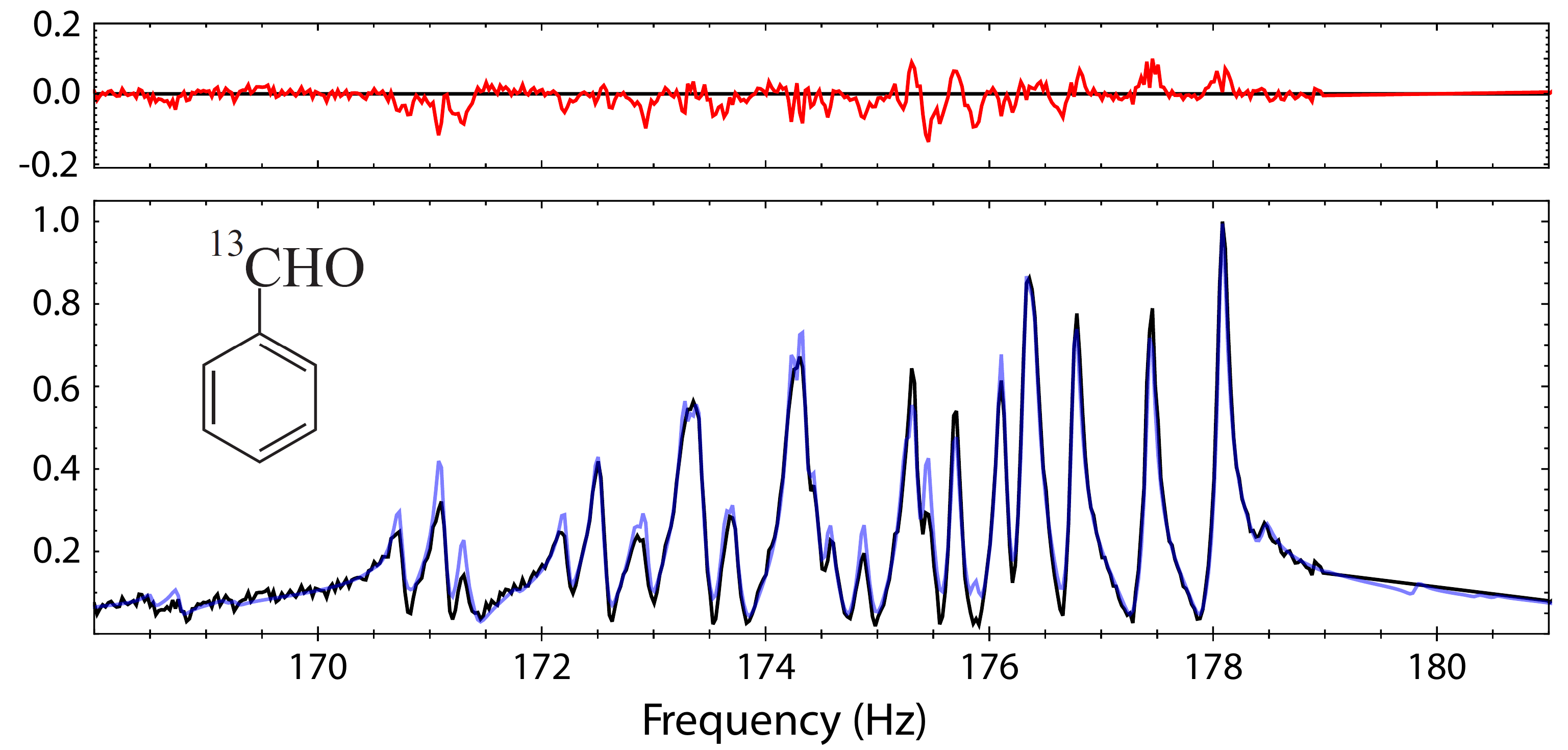}
\caption{Measured zero-field NMR spectrum (black) of benzaldehyde-$\alpha$-${}^{13}$C$_1$ in the vicinity of ${}^1J_{\text{CH}}$ with the fitted spectrum (blue) overlaid. As in Fig.~\ref{Fig:benzene}, noisy low-frequency portions of the experimental spectrum as well as 2 Hz intervals around overtones of the 60 Hz line noise were excluded from the fit. Fit residuals are shown in red above.}
\label{Fig:benzaldehyde}
\end{figure}

The case of {benzaldehyde-$\alpha$-${}^{13}$C$_1$ (experimental spectrum and fit shown in Fig.\ \ref{Fig:benzaldehyde}) is similar to the case of benzene-1-${}^{13}$C$_1$ since the spectra both consist primarily of one relatively narrow frequency range with all the information contained in it.
For toluene-$\alpha$-${}^{13}$C$_1$ (Fig.\ \ref{Fig:toluene}), on the other hand, the spectrum features signals in two distinct regions, centered around ${}^{1}J_{\rm CH}$ and $2\times{}^{1}J_{\rm CH}$: $\sim$120--130~Hz and $\sim$240--260~Hz.
The peaks around ${}^{1}J_{\rm CH}$ arise from transitions between states where the total spin angular momentum, $K=\sum_n I_n$, of the methyl ${}^{1}$H spins is $K=\nicefrac{1}{2}$, and those around $2\times{}^{1}J_{\rm CH}$ arise from transitions between states with $K=\nicefrac{3}{2}$.
We note that the residuals for the two multiplets are slightly biased in opposite directions, due to the frequency dependent sensitivity of the magnetometer.
The data have been calibrated based on a measurement of the phase and amplitude response of the magnetometer to applied frequencies in the interval of 4--400~Hz, however calibration files were, at the time, only prepared on a monthly basis, so the amplitude calibration is less than ideal for these particular data.

The fitted values for the $J$-couplings for benzaldehyde-$\alpha$-${}^{13}$C$_1$ and toluene-$\alpha$-${}^{13}$C$_1$ are shown in Table \ref{tbl:benzaldehydeResults} and Table \ref{tbl:tolueneResults}, respectively.
Discrepancies between the fitted and literature values may be ascribed to the fact that the data for Refs.~\cite{Schaefer1985,Schaefer1989} were acquired for samples diluted in CS$_2$, whereas our data were collected for neat liquid.
Furthermore, similar to the case of benzene, benzaldehyde and toluene possess anisotropic magnetic susceptibilities, so that the molecular alignment occurring in large magnetic fields may also affect high-field NMR measurements of their $J$-couplings. 


\begin{table}[h]
\begin{center}
\begin{tabular}{c*{1}{c}cc}
$J$-coupling & Fitted value (Hz) & Literature value & Difference\\ 
\hline
& & \\
${}^1J_{\rm{CH}}$& 174.839(1) & 174.85$^a$ & 0.011\\
${}^3J_{\rm{CH}}$& 4.882(18) & 4.92$^a$ & 0.038\\ 
${}^4J_{\rm{CH}}$& 0.740(19) & 0.72$^a$ & 0.020\\ 
${}^5J_{\rm{CH}}$& 0.713(25) & 0.69$^a$ & 0.023\\ 
& & \\
${}^4J_{\rm{HH}}$& -0.093(17) & -0.152(2)$^b$ & 0.059(17) \\ 
${}^5J_{\rm{HH}}$& 0.406(19) & 0.431(2)$^b$ & 0.025(19)\\ 
${}^6J_{\rm{HH}}$& -0.063(25) & -0.018(2)$^b$ & 0.045(25)\\ 
& & \\
${}^3J_{\rm{HH}}(\rm H2,H3)$& 7.750(8) & 7.695(2)$^b$ & 0.055(8)\\ 
${}^3J_{\rm{HH}}(\rm H3,H4)$& 7.515(7) & 7.443(2)$^b$ & 0.072(7)\\ 
${}^4J_{\rm{HH}}(\rm H2,H4)$& 1.351(4) & 1.333(2)$^b$ & 0.018(4)\\ 
${}^4J_{\rm{HH}}(\rm H2,H6)$& 1.792(5) & 1.738(2)$^b$ & 0.054(5)\\ 
${}^4J_{\rm{HH}}(\rm H3,H5)$& 1.257(5) & 1.236(2)$^b$ & 0.021(5)\\ 
${}^5J_{\rm{HH}}(\rm H2,H5)$& 0.621(4) & 0.624(1)$^b$ & 0.003(4)\\ 
& & \\
\hline
\end{tabular}
\caption{Fit results for benzaldehyde-$\alpha$-${}^{13}$C$_1$. $^a$From Ref.~\cite{Blanchard2013}. $^b$From Ref.~\cite{Schaefer1989}, as a 4.0 mo1\% solution in CS$_2$ with 10 mol\% C$_6$D$_{12}$ ans 0.5 mol\% tetramethylsilane (TMS).}
\label{tbl:benzaldehydeResults}
\end{center}
\end{table}

\begin{figure*}[!t]
 \includegraphics[width=\textwidth]{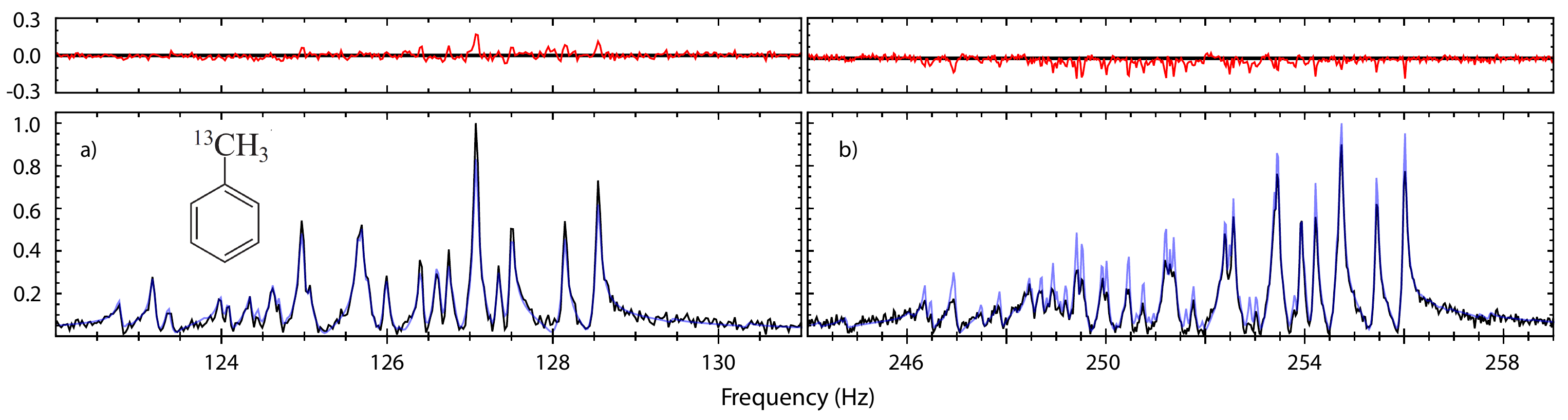}
 \caption{Measured spectrum (black) of toluene-$\alpha$-${}^{13}$C$_1$ in the vicinity of (a) ${}^1J_{\text{CH}}$ and (b) $2\times{}^1J_{\text{CH}}$ with the fitted spectrum (blue) overlaid. The residuals are shown in red above.}
  \label{Fig:toluene}
\end{figure*}

\begin{table}[h]
\begin{center}
\begin{tabular}{c*{1}{c}cc}
$J$-coupling & Fitted value (Hz) & Literature value & Difference\\ 
\hline
& & \\
${}^1J_{\rm CH}$& 125.9895(2) & 125.99$^a$ & 0.0005\\
${}^3J_{\rm CH}$& 4.6053(17) & 4.53$^a$ & 0.0753\\ 
${}^4J_{\rm CH}$& 0.5359(16) & 0.56$^a$ & 0.0241\\ 
${}^5J_{\rm CH}$& 0.6556(23) & 0.63$^a$ & 0.0256\\ 
& & \\
${}^4J_{\rm HH}$& -0.7137(9) & -0.702(1)$^b$ & 0.0135(13)\\ 
${}^5J_{\rm HH}$& 0.3324(14) & 0.329(1)$^b$ & 0.0034(17)\\ 
${}^6J_{\rm HH}$& -0.6020(21) & -0.602(2)$^b$ & 0.0000(21)\\ 
& & \\
${}^3J_{\rm HH}\rm (H2,H3)$& 7.6797(62) & 7.655(2)$^b$ & 0.0247(65)\\ 
${}^3J_{\rm HH}\rm (H3,H4)$& 7.4751(83)& 7.417(2)$^b$ & 0.0581(85)\\ 
${}^4J_{\rm HH}\rm (H2,H4)$& 1.2597(28) & 1.273(3)$^b$ & 0.0133(28)\\ 
${}^4J_{\rm HH}\rm (H2,H6)$& 1.9215(42) & 1.902(3)$^b$ & 0.0195(42)\\ 
${}^4J_{\rm HH}\rm (H3,H5)$& 1.4659(42) & 1.442(2)$^b$ & 0.0239(42)\\ 
${}^5J_{\rm HH}\rm (H2,H5)$& 0.6118(34) & 0.610(1)$^b$ & 0.0018(34)\\ 
& & \\
\hline
\end{tabular}

\caption{Fit results for toluene-$\alpha$-${}^{13}$C$_1$. $^a$From Ref.~\cite{Blanchard2013}. $^b$From Ref.~\cite{Schaefer1985}, as a 2.0 mo1\% solution in CS$_2$ with 10 mol\% C$_6$D$_{12}$ ans 0.5 mol\% TMS.}
\label{tbl:tolueneResults}
\end{center}
\end{table}



\subsection{Formamide-${}^{15}$N}

\begin{figure}[htb]
  \includegraphics[width=0.49\textwidth,height=130px]{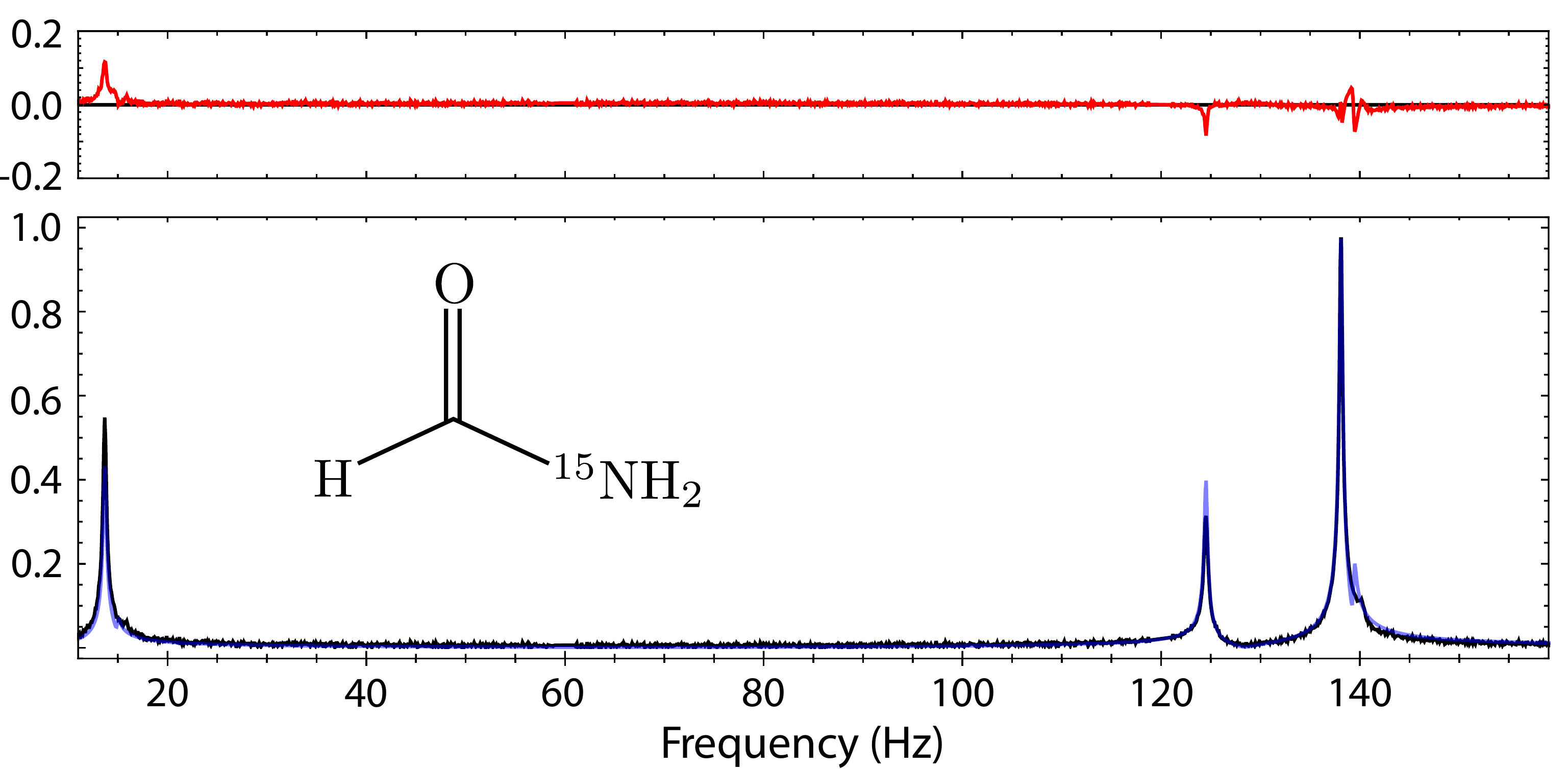}
   \caption{Measured spectrum (black) of formamide-${}^{15}$N in the fast-exchange limit. The fitted spectrum (blue) is overlaid and the residuals are shown in red above.}
  \label{Fig:formamide}
\end{figure}
In addition to its use as a convenient test system while developing the fitting code, formamide-${}^{15}$N is also an interesting example of a spin system that changes based on temperature and measurement field.
Formamide-${}^{15}$N is a relatively simple spin system composed of four spin-1/2 nuclei, with up to six distinct $J$-couplings.
In high-field room-temperature experiments such as described in Ref.~\cite{Sorensen1981F}, it is possible to measure distinct chemical shifts for each nucleus, along with all six couplings.
However, at elevated temperatures, the rate of internal rotation, $k$, increases, eventually causing the two amide protons to become indistinguishable on the timescale defined by the measurement.
In zero-field NMR measurements, inequivalent spins can only be differentiated based on $J$-couplings to other nuclei, and these differences are typically very small compared to the chemical shift differences present in high-field NMR.
For the data presented in Fig.~\ref{Fig:formamide}, we have considered two different models: one in the slow-exchange limit, $k\ll\left|{}^3J_{\rm HH-cis}-{}^3J_{\rm HH-trans}\right|$, and one in the fast-exchange limit, $k\gg\left|{}^3J_{\rm HH-cis}-{}^3J_{\rm HH-trans}\right|$.
In the fast-exchange limit, the spin system is best described as (XA$_2$)B \cite{Blanchard2013,Theis2013}, with three independent $J$-couplings.
In the slow-exchange limit, it is an (XAB)C system, with six independent $J$-couplings.

The fit results for formamide-${}^{15}$N are shown in Tbl.~\ref{tbl:formamideResults}.
Because our uncertainties are estimated based on the diagonal elements of the inverse covariance matrix, it is preferable to choose fit parameters that do not co-vary, so for the slow-exchange limit fit, we defined parameters ${}^1J_{\rm NH}$, $\Delta{}^1J_{\rm NH}$, ${}^3J_{\rm HH}$, and $\Delta{}^3J_{\rm HH}$, such that ${}^1J_{\rm NH-cis}={}^1J_{\rm NH}-\Delta{}^1J_{\rm NH}$, ${}^1J_{\rm NH-trans}={}^1J_{\rm NH}+\Delta{}^1J_{\rm NH}$, ${}^3J_{\rm HH-cis}={}^3J_{\rm HH}-\Delta{}^3J_{\rm HH}$, and ${}^3J_{\rm HH-trans}={}^3J_{\rm HH}+\Delta{}^3J_{\rm HH}$.
As shown in Tbl.~\ref{tbl:formamideResults}, the parameters $\Delta{}^1J_{\rm NH}$, $\Delta{}^3J_{\rm HH}$, and ${}^2J_{\rm HH}$ are zero within error, consistent with the system being in the fast-exchange limit.

\begin{table}[h]
\begin{center}
\begin{tabular}{{r}{c}{c} c}
~ &Fitted value$^a$ (Hz)&Fitted value$^b$ (Hz)& Literature value$^c$\\ 
\hline
& & \\
${}^1J_{\rm NH}$		&-89.335(1)		&-89.335(1) 	&-89.15(14)\\ 
$\Delta{}^1J_{\rm NH}$	&- 				&-0.0005(30) 	&1.35(14)\\ 
${}^2J_{\rm NH}$		&-13.593(2) 	&-13.593(2) 	&-14.25(05)\\ 
${}^3J_{\rm HH}$ 		&7.252(6) 		&7.252(6) 		&7.60\\ 
$\Delta{}^3J_{\rm HH}$ 	&- 				&0.0003(50) 	&5.32\\ 
${}^2J_{\rm HH}$		&- 				&1.5(6.5) 		&2.60\\ 
& & \\
\hline
\end{tabular}
\caption{Fit results for formamide-${}^{15}$N. $^a$In the fast exchange limit. $^b$In the slow exchange limit. $^c$From Ref~\cite{Sorensen1981F}, 90\%v/v formamide in DMSO-$d_6$ at 30$^\circ$C in a $\sim$2.35~T field, the directly measured 1- and 3-bond couplings are ${}^1J_{\rm NH-cis}$=$-87.80(10)$~Hz, ${}^1J_{\rm NH-trans}$=$-90.35(10)$~Hz, ${}^3J_{\rm HH-cis}$=12.92~Hz, and ${}^3J_{\rm HH-trans}$=2.28~Hz.}
\label{tbl:formamideResults}
\end{center}
\end{table}

\subsection{Methyl Formate-${}^{13}$C$_1$}
Methyl formate-${}^{13}$C$_1$ is another relatively simple molecule, consisting of a strongly coupled ${}^{13}$C-${}^{1}$H pair that is in turn coupled to three equivalent methyl protons.
Due to difficulties in obtaining precise literature values for the $J$-couplings for this particular isotopomer, we focus here on the comparison of models with and without additional interaction terms in the Hamiltonian. 
An interesting detail of the methyl formate-${}^{13}$C$_1$ spectrum presented in Fig.~\ref{Fig:methfor} is that for these data, there appears to be some non-Lorentzian broadening of the resonances, such as would arise due to a small residual magnetic field.
%
In \cite{PhysRevLett.107.107601} the effect of small magnetic fields on near-zero-field NMR $J$-spectra is shown to result in additional splittings of the peaks.
While such splittings are not resolved in Fig.~\ref{Fig:methfor}, peculiarities in the lineshapes and the inability to resolve the two peaks at $\sim$225~Hz (resolved in Ref.~\cite{Butler2013}) lead us to believe that a small residual magnetic field may be present. 
By including a longitudinal and transverse field as parameters to the fit, the reduced chi-square 
improves from 1.96 to 0.98 compared to the case without any magnetic field. This is also reflected in the decrease of the standard deviations shown in Table \ref{tbl:methforResults}.
As a result of the fit the transverse magnetic field $B_t$ was estimated to be $\SI{3.1(3)}{\nano\tesla}$ and the longitudinal $B_l$ to be $\SI{2.8(4)}{\nano\tesla}$.

\begin{figure*}[!t]
  \includegraphics[width=\textwidth]{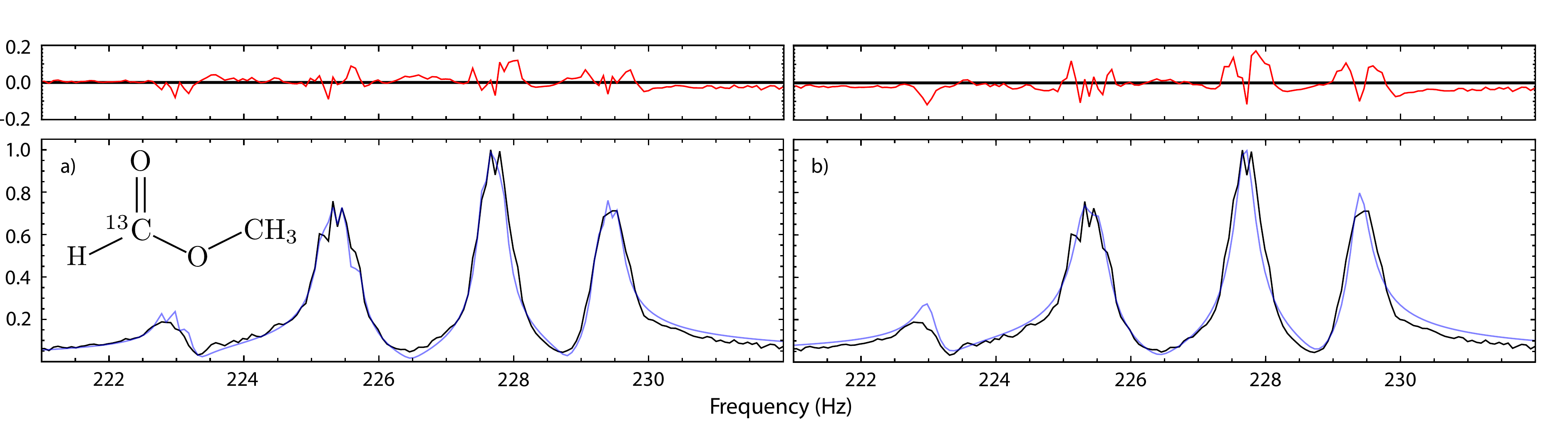}
  \caption{Experimental spectrum, fit, and residuals for methyl formate-${}^{13}$C$_1$ as in Fig.~\ref{Fig:benzene}. Two fits are shown: (a) including a transverse and longitudinal magnetic field, and (b) at zero magnetic field.}
  \label{Fig:methfor}
\end{figure*}

\begin{table}[h]
\begin{center}
\begin{tabular}{c*{1}{c}c}
$J$-coupling & Fitted value$^a$ (Hz) & Fitted value$^b$ (Hz) \\ 
\hline
& & \\
${}^1J_{\rm CH}$& 226.810(3) & 226.792(2)  \\ 
${}^3J_{\rm CH}$& 4.211(40) & 4.267(23)  \\ 
${}^4J_{\rm HH}$& -1.028(40) & -1.068(22)  \\ 
& & \\
\hline
\end{tabular}
\caption{Fit results for methyl formate-${}^{13}$C$_1$. $^a$Assuming no magnetic field. $^b$Including transverse and longitudinal magnetic fields.}
\label{tbl:methforResults}
\end{center}
\end{table}

\subsection{Acetonitrile-${}^{13}$C$_2$,${}^{15}$N$_1$}
Acetonitrile-${}^{13}$C$_2$,${}^{15}$N$_1$ is a special case, in that all of its nuclei have spin 1/2.
As such, there are several strong $J$-couplings, leading to a complex spectrum that is not easily described in terms of a simple strongly coupled XA$_n$ subsystem perturbed by additional weaker long-range couplings, as in Refs.~\cite{Blanchard2013,Theis2013,Butler2013}.
The result is that the spectrum contains peaks across a comparatively wide range, as can be seen in Fig.\ref{Fig:acetonitrileCorr_Abs}.
While such a range of resonances may be useful for broadband experiments searching for oscillatory exotic interactions (e.g. Refs.~\cite{PhysRevD.88.035023, CASPEr2014}), the fit is susceptible to the frequency-dependent sensitivity and phase response of the magnetometer.
The data presented in Fig.\ref{Fig:acetonitrileCorr_Abs} were thus calibrated using the same phase and amplitude response measurements as were used above for toluene-$\alpha$-${}^{13}$C$_1$.
The resulting measured couplings are reported in Table \ref{tbl:acetonitrileResults}, and are in good agreement with the values reported in Refs.~\cite{SORENSEN1981,DIEHL1982}, with deviations due to solvent effects perhaps even smaller than expected compared to solvent dependences noted in Ref.~\cite{DIEHL1982}.
%
%
%
%
%
\begin{figure}
 \includegraphics[width=0.49\textwidth,height=130px]{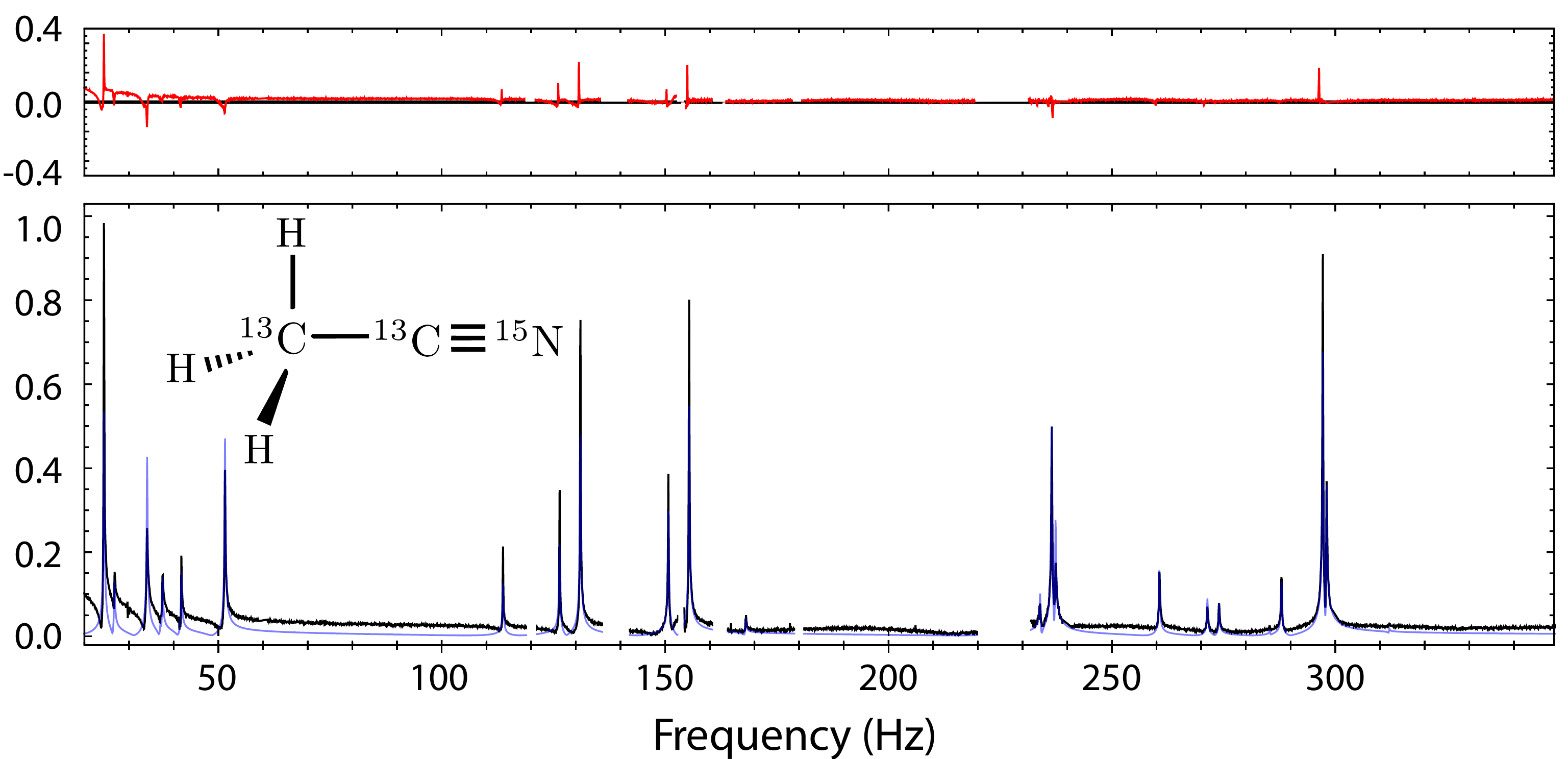}
  \caption{Measured zero-field NMR spectrum (black) of acetonitrile-${}^{13}$C$_2$,${}^{15}$N$_1$ with the fitted spectrum (blue) overlaid. Regions at $\sim$140~Hz, $\sim$160~Hz, and $\sim$225~Hz with a slightly increased noise floor (but isolated from spectral resonances), along with 2 Hz intervals around overtones of the 60 Hz line noise were excluded from the fit. Fit residuals are shown in red above.}
\label{Fig:acetonitrileCorr_Abs}
\end{figure}
\begin{table}
\begin{center}
\begin{tabular}{c*{1}{c}cc}
$J$-coupling & Fitted value (Hz) & Literature value & Difference\\ 
\hline
& & \\
${}^1J_{\rm CH}$& 136.200(1) & 136.25(10)$^a$ & 0.050(100)\\ 
${}^2J_{\rm CH}$& -9.924(2) & -9.94(4)$^a$  & 0.016(40)\\ 
${}^3J_{\rm NH}$& -1.688(5) & -1.69(2)$^a$ & 0.002(21)\\ 
${}^1J_{\rm CC}$& 57.010(5) & 56.94(4)$^a$ & 0.070(40)\\ 
${}^2J_{\rm CN}$& 2.822(5) & 2.9(2)$^b$ & 0.078(200)\\ 
${}^1J_{\rm CN}$& -17.419(5) & -17.53(10)$^a$ & 0.134(100)\\ 
& & \\
\hline
\end{tabular}
\caption{Fit results for acetonitrile-${}^{13}$C$_2$,${}^{15}$N$_1$. $^a$From Ref.~\cite{SORENSEN1981}, 90\%v/v aceonitrile in acetone-$d_6$, $^b$From Ref.~\cite{DIEHL1982}, in CDCl$_3$.}
\label{tbl:acetonitrileResults}.
\end{center}
\end{table}
\subsection{Systematic Effects}
Considering systematic effects on $J$-couplings one of the main concerns is spatial and temporal temperature gradients. 
Since certain temperatures allow transitions between different electronic structural modes, the coupling constants can be influenced \cite{doi:10.1021/ja00149a025}.  
The fact that all spectra shown here consist of averaged signals which were taken over a longer period, usually over several hours, required a test to determine time dependent systematic biasing of the data.
By analyzing every individual spectrum of methyl formate-${}^{13}$C$_1$, formamide-${}^{15}$N and benzaldehyde-$\alpha$-${}^{13}$C$_1$ the fitted values of every parameter show a normal distribution and no time dependence is observed, meaning that the apparatus is not, for example, changing the sample's temperature in a observable manner over long timescales.

There are also time-independent effects like sample impurities or residual external magnetic fields.
%
%
However, as shown for methyl formate, external magnetic fields can be included as parameters to the procedure, so that those can be incorporated in the fits.
Furthermore, in a situation where the interest is more on the variation of the $J$-couplings in respect to a specific property, with otherwise constant external conditions, ``static" systematic effects become less important. 

The presented method does not account for finite-pulse effects since all spins are tipped instantaneously in the simulation.
The primary consequence of using finite-length pulses in ZULF NMR experiments is the presence of small phase differences as compared to the delta function limit.
For this reason the absolute value of the spectrum is fitted rather than the phase-dependent real and imaginary parts. 

\section{Conclusions}
Based on numerical simulation and least-squares fitting, estimates of $J$-coupling constants have been extracted from ZULF spectra with a precision of $10^{-2}$ to $10^{-4}$ Hz.
The majority of the extracted $J$-coupling values are in agreement with earlier results and are of higher precision.

The simulation and fitting routine can easily be modified to incorporate additional interaction terms, time-dependent interactions, and complex pulse sequences.
The method can readily be used to measure $J$-couplings of molecules which have not been analyzed before.
Precise $J$-coupling determinations can also be used to study fundamental interactions. 
For example, the presence of exotic spin-dependent forces such as those described in Ref.~\cite{Dobrescu2006} would be expected to manifest as additional terms in the Hamiltonian.
It might also be possible to search for symmetry breaking from molecular parity non-conservation.
Inclusion of time-dependent Hamiltonians may also allow for searches for oscillating fields associated with dark matter (axions or axion-like particles) as proposed in \cite{PhysRevD.88.035023}.

\section{Acknowledgment}
This work was supported by the National Science Foundation under award CHE-1308381, 
the DFG Koselleck program, and the Heising-Simons and Simons
Foundations. 
This project has also received funding from the European Research Council (ERC) under the European
Union's Horizon 2020 research and innovation programme (grant agreement No 695405).

\bibliographystyle{apsrev4-1.bst}
\bibliography{ZULF-fitting}

\end{document}